\documentclass[12pt,thmsa]{article}
\usepackage{amssymb}

\input{tcilatex}
\begin{document}

\author{V.S. Eganov$^{(*)}$,A.P. Garyaka$^{(*)}$, E.V. Korkotian$^{(*)}$, \\
%EndAName
E.A. Mamidjanian$^{(**)}$, R.M Martirosov$^{1(*)}$, J. Procureur$^{(***)}$, 
\\
H.E. Sogoyan$^{(*)}$, M.Z. Zazyan$^{(*)}$\\
\\
\vspace*{-3mm} {\footnotesize $(*)$ Yerevan Physics Institute, Br. Alikanian
St.2, 375036 Yerevan, Republic of Armenia}\\
\vspace*{-3mm} {\footnotesize $(\ast\ast)$ P.N. Lebedev Institute Leninsky
pr. 56, Moscow 117924, Russia}\\
\vspace*{-3mm} {\footnotesize $(***)$ Centre d Etudes Nucleaires de
Bordeaux-Gradignan, Universit\'{e} de Bordeaux 1}\\
\vspace*{-3mm} {\footnotesize rue du Solarium, 33175 Gradignan-Cedex, France}}
\title{Analysis of the experimental results from the EAS installation GAMMA at Mt.
Aragats}
\maketitle

\begin{abstract}
The phenomenological characterstics of the electron and muon components of
EAS with the size $10^{5}\leq N_{e}\leq 10^{7}$ are obtained by the GAMMA
installation of the ANI experiment at the Mt. Aragats in Armenia at the
observation level 700 $g.cm^{-2}$. The experimental results are compared
with other experiments and with the simulation carried out using the CORSIKA
code.
\end{abstract}

\section{\protect\footnote{%
Corresponding author. E-mail: martir@lx2.yerphi.am}Introduction}

The particle physics and astrophysics aspects of air shower studies are
closely interlaced. The problems of sources and nature of the primary
radiation and features of the high energy hadron interactions cannot be
solved separately. The history of attempts to clarify the origin of the
appreciable change in the index of the shower size spectrum (the $knee)$,
which was found more than 40 years ago \cite{Kulikov} and confirmed by other
experiments, is the best manifestation of that. Many new experiments were
born devoted to the observation of various components of EAS, including
Cherenkov light in the atmosphere and experiments under water. However, up
to now there is no unequivocal explanation of the reason of this phenomena.
The most popular interpretation of the $knee$ is the existence of the $knee$
in the primary cosmic ray energy spectrum at energy about (3-5)$^{.}$10$^{%
\text{6}}$ GeV and, as a possible consequence, the change of the mass
composition. There are different experimental analysis leading to opposite
conclusions about primary composition after $knee$ (lighter \cite{Blake},
normal \cite{Fomin}\cite{Rhode} or heavier \cite{Roth}\cite{Engler}\cite
{Glasmacher}).

At the same time there is an alternative explanation of the $knee$, also
based on some inconsistencies in the EAS data, connected with a modification
in the hadronic interaction properties \cite{Nikolsky1}\cite{Nikolsky2} to
the change of the number and energy spectrum of the secondaries. The complex
installations, which measure electromagnetic component as well as muon and
hadron components of EAS have the best opportunities to explore the nature
of the $knee\mathbf{.}$\ HADRON-2 \cite{Adamov} on Tien-Shan and KASCADE 
\cite{Klages} are installations of such a kind. The hadron component data
obtained from these installations show the fast absorption of the high
energy hadrons, which is impossible to be explained by the energy increase
of the interaction cross-section only. At the same time, the electromagnetic
component data are in an agreement with the QGSJET model and correspond to
the mixed composition of primary with progressive changing of the mass
composition. In the work \cite{Nikolsky2} on the bases of the Tien-Shan data
analysis it is supposed that there is no $knee$ in the primary spectrum and $%
knee$ in the EAS size spectrum is explained by the change of the hadronic
interaction in the atmosphere. This explanation is based on the results of
the EAS electromagnetic component study \cite{Vildanova}, according to which
there is no $knee$ neither in the $young$\ showers spectrum (showers with $%
S<0.75,$ generated by primary protons interacted in the depth of
atmosphere), nor in the $old$ showers ($S>1.05$) one. And the existence of
the $knee$\ in the same spectra of other experiments could be explained by
the missing of the young showers.

On the other hand it is not so clear situation with the so-called $reverse$ $%
knee$ in the EAS size spectrum at $N_{e}\thicksim (2-3)*10^{7}$, which was
discovered in \cite{Fomin1}. Subsequently the same irregularity was obtained
once again in Tien-Shan experiment \cite{Vildanova}.

At present, practically all experiments on research of the cosmic rays in
the energy range 10$^{14}$-10$^{17}$\ eV are pointed to investigate the
nature of the $knee$. We have to point out that the KASCADE \cite{Klages},
CASA-MIA \cite{Aglietta} and EAS-TOP \cite{Barione} experiments present the
basic advantage to have their data analyzed using multiparameter procedures.
Only such an approach to study the EAS characteristics allows some advance
in understanding of the reasons of the existence of the $knee$ in the
spectrum on N$_{e}$. In this paper, we present the experimental results of
the GAMMA installation of the experimental complex ANI also working in this
field.

The GAMMA installation is located on hillside of the Mt. Aragats in Armenia
(3200 m a.s.l.). In comparison with the other large scale experiments with
the same goals, the high-altitude observation level of the GAMMA
installation provides some advantages, in particular for minimizing the
intrinsic fluctuations of the observables due to the stochastic character of
the EAS development in atmosphere. This will add some new observations,
which would clearly complement the data from KASCADE (near sea-level),
CASA-MIA (1200 meters), and EAS-TOP (2000 meters). In the present work,
showers are simulated using the CORSIKA code 5.20 \cite{Heck} in which the
hadronic interactions are described by the VENUS model,\cite{Werner}. On the
other hand, to compare the simulation data with experimental ones, the
normal mixed composition was used, i.e. proton-40\thinspace $\%$, $\alpha $%
-21$\%$, light-nuclei (\TEXTsymbol{<}A\TEXTsymbol{>}=14)-14$\%$,
medium-nuclei (\TEXTsymbol{<}A\TEXTsymbol{>}=26)-13$\%$\ and heavy-nuclei (%
\TEXTsymbol{<}A\TEXTsymbol{>}=56)-12$\%$

\smallskip

\section{Present status of the GAMMA EAS installation}

The GAMMA installation \cite{Arzumanian} was realized as a part of the
project ANI \cite{Danilova} in an attempt to continue experimental study of
the hadronic interactions and the energy spectrum and mass composition of
the primary cosmic radiation in the energy range of $10^{14}-10^{17}$ eV.

After some years spent to enlarge the effective area of the muon underground
detector, which was necessary to elaborate methodical studies of the
detector parameters, to investigate carefully the array response and to
determine the precision of the shower parameter estimation, the GAMMA
experiment is, now, effectively running.

GAMMA is a central type array and consists of two main parts $(figure\,1)$:

(i) the surface part, for the registration of the EAS electromagnetic
component;

(ii) the large muon underground detector, to register the EAS muon component.

\smallskip

\subsubsection{The surface scintillation array}

The surface scintillation array constists of 25 groups of 3 scintillation
detectors placed on concentric circles with radii of 17, 28, 50 and 70 $m$.
Each of the detectors has 1 $m^{2}$ of effective area. They are distributed
on the full area of $\thickapprox $1.5$\cdot 10^{4}$ $m^{2}$. An additional
station with 20 detectors of the same type is placed at the 135 $m$ from the
installation centre. The energy threshold for the registration of the
electrons flux, $E_{e}^{thres}\simeq 9.5$ MeV, is estimated taking into
account the thickness and the specificities of the detectors and
registration stations. The calibration of detectors has been carried out by
periodical (each hour) registration of the cosmic ray background particles .
It was taken into account that the most probable value of the registered
particles differs from $0^{0}$ and depends on the observation level \cite
{Mnatsakanyan}$.$

Each of the 25 registration stations is equipped with the timing channel
which allows to determine the angular coordinates of the shower axis. The
trigger condition for one EAS registration is that in each of four groups of
detectors placed at 17 meters from the array centre, two detectors have to
register a flux density with $\rho \geq 3\,\,part/m^{2}$.

The main EAS parameters: coordinates and angles of the shower axis $%
(X,Y,\theta ,\varphi )$, number of electrons $(N_{e})$ and age parameter $%
S_{NKG}$ are obtained using the SPACE code \cite{Aseikin} created on the
basis of the statistical methods for the solution of the inverse problem 
\cite{Pavlyuchenko}. Taking into account the detector response, GAMMA array
geometry and using Monte-Carlo procedure the data bank of pseudo
experimental showers was created and treated by the same SPACE code.

As the result of this study, the following accuracies on the EAS parameter
estimations are obtained: $\Delta X,\Delta Y\leqslant 3m$ $($for $R\leq 40m)$%
, $\sigma _{\theta }\backsimeq 1.5^{0}$, $\sigma _{\varphi }\backsimeq 8^{0}$%
, $\sigma _{N_{e}}/<N_{e}>\leq 20\%$.

Furthermore, the analysis of the shower registration efficiency shows that
the showers with particle number $N_{e}\gtrsim 3\cdot 10^{5}$ within 20 $m$
and $N_{e}\gtrsim 10^{6}$ within 40 $m$ from the array centre are registered
with 100\% efficiency.

\smallskip

\subsubsection{Underground muon detector}

The layout of the muon underground detector is shown in $figure\,2$. As for
the GAMMA surface part identical scintillation detectors with $S=1m^{2}$
were used for the registration of the EAS muon component. It should be noted
that the muon scintillation detectors are divided into 2 groups placed under
different absorber thickness. This allows to study EAS muons with 2
different energy thresholds: $E_{\mu }\gtrsim 5\,$GeV (Hall) and $E_{\mu
}\gtrsim 2.5\,$GeV (Tunnel). These thresholds were estimated experimentally
using scintillation telescope with a small solid angle. The arrangement of
the muon detectors gives the possibility to determine the muon lateral
distribution up to $60m$ at $E_{\mu }\gtrsim $ 5$\,$GeV and up to $90m$ at $%
E_{\mu }\gtrsim \,$2.5$\,$GeV.

Some peculiarities of the GAMMA array geometry make necessary to pay
attention on the azimuth angle symmetry of data. The surface detectors are
placed on the hillside and sometimes the difference between $Z$ coordinates
of the individual detectors is reaching $18m$. On the other hand, the
position of muon detector is very asymmetric in regard to the geometric
centre of our installation. In this case, the influence of azimuth angle $%
\varphi $ could be strong. To check this effect we divided the observed
showers into four groups with $\varphi \in
[0-90]^{0},[90-180]^{0},[180-270]^{0},[270-360]^{0}$. The comparison of the
average muon lateral distributions for each of these groups by $\varphi $ is
shown in $figure\,3\,(a,b)$ for $<N_{e}>=1.28\cdot 10^{6}$. It can be
visible, that the difference of the muon lateral distributions for the
different ranges of $\varphi $ is quite negligible. Furthermore, it is
necessary to point out that for the small distances (less than $15\,m$ for $%
E_{\mu }\gtrsim $2.5 GeV and $8\,m$ for $E_{\mu }\gtrsim $5 GeV) there is a
significant influence of the punch through of particles (high energy
electrons and hadrons) and these distances are excluded from the further
analysis.

\strut

\section{Results}

We have used the experimental data obtained during 3300 hours operation
time. The number of EAS with $N_{e}\geq 10^{5}$ and with zenith angle $%
\theta \leq 30^{0}$ is $\backsim 260.000$. The effective area for the
selection of the EAS axis is $\backsim 5.000m^{2}$.

\smallskip

\subsection{Electromagnetic component characteristics}

The main parameters of the EAS electromagnetic component at the given
observation level are: the total number of the charge particles or shower
size $N_{e}$ and the lateral shower age parameter $S$. In order to obtain
the EAS size spectrum and to investigate its behaviour in the $knee$ region
the correct estimation of these parameters is needed.

\smallskip Usually in the energy region of $10^{5}-10^{7}\,$GeVthe
scintillation detectors are used to register the EAS charged particles.
However it is well known that such type of a detector registers not only the
EAS charged particles, but also electrons generated in the absorber above as
well as in the scintillator by EAS photons. This contribution to the
measured charged particle density depends both on the scintillator thickness
and the distance of detector from the shower axis.

\smallskip The comparison of the charge particle densities seen by
scintillation detectors and Geiger counters has been made for different
distances from the shower core \cite{Wallace}. In the same way, comparisons
with other kind of detectors have been performed, (spark chamber \cite
{Shibata}, Geiger counter \cite{Chudakov}, neon flash tube \cite{Bagge}, as
well as with thin and thick scintillators \cite{Hara}\cite{Asakimori}). But,
up to now, there are not exact estimation of the photon contribution. In the
energy range $10^{5}-10^{7}\,\,$GeV\ all these investigations show both a
smooth rise and a noticeable decrease of the photon contribution to the
measured shower size versus the distance to the shower core up to $100\,m$.

Recently the methodical experiment on the MAKET-ANI installation on the Mt.
Aragats has been carried out \cite{Blokhin}.\textbf{\ }Comparing the
densities measured by scintillators with different thicknesses ($1.0,\,1.5$\
and $5.0\,cm$), up to $100\,m$, the ratio $K_{\gamma }(r)$\ of the measured
density $\rho _{sc}(r)\,$to the charged one $\rho _{ch}(r)\,\,$is described
by $K_{\gamma }(r)=\rho _{sc}(r)/\rho _{ch}(r)=(r/R_{M})^{-0.18}$, where $r$%
\ is the distance from the shower core and $R_{M}=120\,m$\ is the Moliere
radius for the observation level of $700\,g.cm^{2}$.\textbf{\ }According to
this result the photon contribution to the charged particle density is $%
\backsim 17\%$\ at $50\,m$\ from the shower axis and practically disappears
at $100\,m$.

At the same time, simulations show \cite{Weber}\cite{Sanosyan}\ a
significant increase of the ratio of the photon number to electrons one
versus $r$. Calculations for the real conditions of the GAMMA installation
have been performed taking into account the absorber and scintillator
thicknesses \cite{TerAntonian}. According to this work the photon
contribution to the measured number of the charged particles is practically
constant from $5$\ to $100\,m$\ and is $K_{\gamma }\backsimeq 1.25$. The
processing of the GAMMA data considering these two results, $K_{\gamma
}(r)=(r/R_{M})^{-0.18}$ and $K_{\gamma }=1.25$,\cite{Blokhin}\cite
{TerAntonian}$\,$, has been made to study the influence of this factor on
the measured EAS features.

In $figures$ $4\,$ the experimental electron lateral distributions are shown
for three EAS average size intervals $<N_{e}>$=$2.3\,*10^{5},$ $7.3\,*10^{5}$
and $22.8\,*10^{5}$ as estimated for the two cases of the photon
contribution. It can be seen that in both cases the electron lateral
distribution can be very well approximated using the NKG function, but
because of the different shape of $K_{\gamma }$\ versus $r$, the average age
parameter value $<S>$\ is different by about 0.16$.$

The comparison between the measured and simulated electron lateral
distributions is shown in $figure\,5$\ for the same values of $<N_{e}>$. It
can be seen a good agreement for the correction factor $K_{\gamma
}=1.25=const$. This agreement improves with the increasing of $<N_{e}>$.

It has to be noticed that the age parameter only describes the shape of the
charged particles lateral distribution and its values $<S>$ $<1$ do not
contradict to the fact that the shower is over the longitudinal development
maximum. $Figure$\ $6$\ shows the dependence of the age parameter $<S>$\
versus the shower size $<N_{e}>$ from our experimental data processed with $%
K_{\gamma }(r)=(r/R_{M})^{-0.18}$ and $K_{\gamma }=1.25$ and the results
from our simulation and the Tien-Shan experimental data \cite{Adamov1}. For $%
<N_{e}>$ $\geq $\ $(3-5)\,*10^{5}$, experimental data with $K_{\gamma }=1.25$%
\ are in a good agreement with the simulation results which is not at all
the same in the case of $K_{\gamma }(r)=(r/R_{M})^{-0.18}$\ for which the
disagreement is large.

\smallskip For the both variants of processing, the differential size
spectra are shown in $figure$ $7a$ for showers in the zenith angle interval $%
\theta \leqslant 30^{0}$, (mean atmospheric depth : $738\,g.cm^{-2}$). In
both cases the $knee$ can be easily observed. Spectra are approximated in
the whole $N_{e}$ interval by formula \cite{TerAntonian}: $%
I(N_{e})=A*N_{e}^{-\gamma _{1}}(1+(N_{e}/N_{e}^{knee})^{k})^{-(\gamma
_{1}-\gamma _{2})/k}$. The spectra differ mainly at the beginning and the
end of the size interval, (by no more than 20\%). Spectral indexes
difference for GAMMA for both cases is $\Delta \gamma =0.36\pm 0.02$. The $%
knee$ regions approximately coincide at $N_{e}^{knee}\thickapprox
(1.8-2.0)*10^{6}$. The integral intensity for the range of $N_{e}\geqslant
N_{e}^{knee}$ is $(1.3\pm 0.4)\,\,10^{-7}m^{-2}s^{-1}sr^{-1}$. This result
is in agreement with the data of other experiments \cite{Navarra}\cite{Rebel}%
.

In order to demonstrate the $knee$ region properties more visually spectra
are given in $figure$ $7b$ multiplied by $N_{e}^{2.5}$. For the case of $%
K_{\gamma }=1.25$ spectral index is less by $(0.07\pm 0.02)$ in the whole $%
N_{e}$ region.

In the $figure$ $8,$ the size spectra for the two zenith angle intervals are
shown at $<\theta >=14^{0}$ and $<\theta >=31^{0}$ to make the qualitative
assessment of the behaviour of the spectra at different angles. They are
practically parallel and coincide when data with $<\theta >=31^{0}$ are
shifted. Such behaviour corresponds to the shower absorption lenght of $%
\Lambda =(230\pm 25)g\cdot cm^{-2}$\textbf{\ }below the knee and $(230\pm
40)g\cdot cm^{-2}$\ above up to $Ne=5\cdot 10^{6}$. The permanency of the
attenuation length is one evidence of the constant charged particle
composition of the showers below and above the $knee$\ only and says nothing
about the cause of the $knee$.

In $figure$ $9$ the spectra are shown with various cuts of the age parameter 
$S$. Relatively $old$ shower spectrum $(S>0.85)$ practically has no $knee$
and is steeper than all shower spectrum. On the other hand, the $young$
shower spectrum $(S<0.75)$ has the obvious $knee$ and is flatter, than the
all shower spectrum below the $knee$ and is almost parallel to it above.
These spectra may be explained with both hypothesises of the $knee$ origin
noted in the introduction, because of the region of $N_{e}\geq (1-2)\cdot
10^{6}$ is transitional.\ The $young$ showers fraction does not arise above
the $knee$ because of change of the mass composition to heavier one in the
model of primary $knee$. In the case of change of the hadron interaction the
fraction of the heavy-like events will rise among the showers generated by
primary protons.

\subsection{Muon component characteristics}

\smallskip As it was mentioned above, the underground muon detectors are
placed in the hall and tunnel of the GAMMA underground part. Due to the
different shieldings of concrete, iron and ground above detectors, muons
with two energy thresholds $E_{\mu }\geqslant 5\,$GeV (hall) and $E_{\mu
}\geqslant 2.5\,$GeV (tunnel) are detected. The muon lateral distributions
for given thresholds and three $<N_{e}>$\ intervals are shown in $%
figure\,10\,a$. These distributions are limited by the number of detectors
and their disposition in the hall and tunnel. Moreover, close to the shower
core the signal in the detectors are strongly affected by the contribution
of other type of particles due to the punch through effect. So, the correct
determination of the muon lateral distribution is available at distances\ $%
\mathbf{[}8$ $-$ $52]\,m$ for $E_{\mu }\geqslant 5\,$GeV and $[20-90]\,m$
for $E_{\mu }\geqslant 2.5\,\,$GeV. In the interval of $[0-50]\,m$, the
shapes of lateral distribution are very similar for the both muon energy
thresholds. This gives the possibility to find a transition coefficient from
the density of muons with $E_{\mu }\geqslant 2.5\,$GeV to the muon density
with $E_{\mu }\geqslant 5\,$GeV, which is $\rho (2.5\,$GeV$)/\rho (5\,$GeV$%
)\approx 1.35$, for all $<N_{e}$ $>$ intervals. In $figure$ $10\,b$, the
muon lateral distribution with the energy threshold $E_{\mu }\geqslant 5\,$%
GeV for the hall detectors and reconstructed one from the distribution of
tunnel detectors are presented. These distributions are well approximated by
the Hillas function, \cite{Hillas}, $\rho _{\mu
}(N_{e},r)=0.9\,(N_{e}/10^{5})\,\,r^{-(0.75+0.06\cdot \ln
(N_{e}/10^{5})}\,\exp (-r/80).$ This formula describes well the experimental
data for $N_{e}<2.10^{6}$. Farther, the dependence of $\rho _{\mu }$ versus $%
N_{e}\,$becomes stronger but the shape of the distribution remains the same.

The approximations of the Tien-Shan data \cite{Aseikin1}$,$ for the same
energy threshold $E_{\mu }\geqslant 5\,$GeV and the same observation level $%
700\,g\cdot cm^{-2}$ are presented by $\rho _{\mu
}(N_{e},r)=0.95\,\,(N_{e}/10^{5})^{0.8}\,\,r^{-0.75}\,\exp (-r/80)$. This
formula describes the GAMMA experimental data noticeably worse, especially
for $N_{e}>10^{6}$.

As it was mentioned above, the lateral distribution of muons in our
experiment,$(E_{\mu }\geqslant 5\,$GeV), is studied for distances $%
r\leqslant 52\,m$ from the shower core where about $33\%$ of all muons are
contained. Obtaining the total muon number, $N_{\mu },$by integration the
muon lateral distribution function, the possible extrapolation errors for $%
r>52\,m$\ will give an error in the $N_{\mu }$ estimation.\ On the other
hand, closer to the shower axis, $r\lesssim 8\,m,$\ other kinds of particles
give a considerable contribution to muon lateral distribution. For this
reason, we use the truncated muon number, $N_{\mu }^{trun}$, which is the
number of muons in the ring $8\,m<r<52\,m$. Experimentally, the truncated
muon number is determined as $N_{\mu }^{trun}(r)=\sum (\rho _{i}^{ex}/w_{\mu
}^{trun}(r))\,/\,K$ , where:

$K$ is number of detectors in the given interval of $r$;

$\rho _{i}^{ex}$ is the experimental muon density in the $i^{th}$ detector
and

$w_{\mu }^{trun}(r)$ is the truncated muon probability distribution,
determined by our muon lateral distribution approximation.

\textit{\smallskip }It should be noticed, that according to simulations, $%
N_{\mu }^{trun}$(\TEXTsymbol{>}5\thinspace GeV)\ in our experiment\ depends
on the primary particle mass. $Figure$\ $11$\ shows $N_{\mu }^{trun}$\
dependence on $N_{e}$. Up to $N_{e}\thickapprox 2\cdot 10^{6}$\ it can be
fitted with the expression $N_{\mu }^{trun}\backsim N_{e}^{0.79}$. For
larger $N_{e}$\ dependence becomes steeper.\textit{\ }It should be noted
that because of the shower size threshold $N_{e}>10^{5}$ the registration of
the shower with fixed muon size and 100\% efficiency is possible at $N_{\mu
}^{trun}\gtrsim 10^{3}$.

The differential $N_{\mu }^{trun}$ spectrum is plotted on $figure$ $12$. It
can be seen that there is no obvious $knee$, but the data do not contradict
to change of spectral index by (0.1-0.2) at $N_{\mu }^{trun}>5\cdot 10^{3}$%
.\ Contrary to the case of fixed $N_{e}$ the showers at fixed $N_{\mu }$ are
enriched by the showers generated by heavy primaries. If the proton spectrum
has a $knee$ only then the $N_{\mu }$ spectrum will have feebly marked $knee$%
. \textbf{\ }At energies above the possible $knee$ position of the primary
iron spectrum, $N_{\mu }$\ spectrum will obtain the final index different by 
$\Delta \gamma \approx 0.5$\ from present one. It is necessary to expand our
measurements to this energy region.

\smallskip Situation in the case of the change of hadron interactions for
sufficientely high energies depends on details of the secondary spectra but
the difference of indexes will be not bigger.

In order to interpret our experimental results the instrumental response has
to be taken into account. For this purpose the detector simulation program
ARES (ARagats Events Simulation) \cite{Haungs} based on the GEANT package.%
\cite{GEANT} has been developed for the GAMMA installation. The data from
CORSIKA simulated EAS are used as input for the ARES code.

To take into account the detector response to EAS muon component all
secondary particles of EAS at the ground level are passed through the
absorber and muon detectors. For each shower the deposited energy and the
muon number in the individual detector are obtained (the methodical
procedure of muon number estimation is described in \cite{Zazyan1}\cite
{Zazyan2}). The muon lateral distribution are derived using the CORSIKA
simulation data in the primary energy range of 3*10$^{5}$-10$^{7}$GeV for
the different primary groups. $Figure$ $13\,(a,b)$ shows the muon lateral
distributions in the shower size range 1.78$\cdot $10$^{5}$\TEXTsymbol{<}N$%
_{e}$\TEXTsymbol{<}3.16$\cdot $10$^{5}$ in the case of the normal mixed
composition for the tunnel and hall detectors. The experimental distribution
and the CORSIKA/ARES simulation results (with the corresponding muon
thresholds) are in a good agreement.

\section{Conclusions}

The study of the EAS electrons and muons $(E_{\mu }>5\,$GeV and $E_{\mu
}>2.5\,$GeV$)$ by the GAMMA array gives reliable information about EAS
characteristics in the shower size range of 10$^{5}$-10$^{7}$. There are not
noticeable contradictions of our data with other experiments. Using the
contribution of the EAS gamma-quanta equal $K_{\gamma }=1.25=const$ an
agreement between CORSIKA simulation and experimental data is obtained.

Permanency of the age parameter $S$ at $N_{e}>2\cdot 10^{6}$ and steepening
of $N_{\mu }^{trun}$ depending on $N_{e}$ at the same range by $N_{e}$
conform to the more rapid shower development. The data in the shower size
interval $N_{e}^{knee}\div 10\cdot N_{e}^{knee}$\ are not decisive for the
problem of the $knee$ origin and enlargement of the study range is necessary
for the complex installations to make sure conclusion about the primary
composition and spectra. With the object to extend a possibilities of the
GAMMA installation it is expected to decrease the registration threshold by $%
N_{e}$ and to enlarge an effective area of the shower selection with $%
N_{e}>10^{6}.$

\smallskip The present paper is based on the ANI collaboration data bank and
express the point of view of the given group of co-authors.

\subsection{Acknowledgments}

The present investigations are embedded in a collaboration between the
Moscow Lebedev Institute (Russia), the Yerevan Physics Institute (Armenia)
and the Universite de Bordeaux 1 (France).

We give thanks to the Staff of the Aragats Research Station for the
assistance during the longterm maintenance of the GAMMA installation.

We are grateful to all colleagues of the Moscow Lebedev Institute and
Yerevan Physics Institute Cosmic Ray Divisions who were taking part in the
development and creation of the GAMMA installation. We would like to express
our special gratitude to prof. S.I. Nikolsky for very useful comments and
N.M. Nikolskaya for the creation of the software.

We thank also prof. A.A.\ Chilingarian, prof. S. Ter-Antonyan and Dr.
E.Mnatsakanyan for discussions and we do not forget P. Aguer from the CENBG,
(CNRS-In2p3-France) whose help has been useful.

This work was supported by the grant 96-752 of the Armenian Ministry of
Industry, by the Russian RFBR 96-02-18098 grant and by the Russian Energy
Ministry.

\section{\protect\smallskip Figures}

Figure 1. Layout of the GAMMA installation

Figure 2. Schematic view of the muon underground detector

Figure 3. Muon lateral distributions at the different azimuth angles $%
\varphi $:

a) hall, b) tunnel

Figure 4. Electron lateral distributions for two cases of the photon
contribution

a) $K_{\gamma }(r)=(r/R_{M})^{-0.18}$, b) $K_{\gamma }(r)=1.25=const$

Figure 5. Electron lateral distribution in comparison with the CORSIKA
simulation for two cases of the photon contribution

Figure 6. The average age parameter $<S>$ versus shower size $<N_{e}>$ at
two cases of the photon contribution in comparison with CORSIKA simulation
and Tien-Shan data

Figure 7. Differential size spectra at two cases of the photon contribution

Figure 8. Differential size spectra at different zenith angles

Figure 9. Differential size spectra at different interval by age parameter

Figure 10. Muon lateral distribution at different shower sizes in

a) hall, b) tunnel

Figure 11. Average truncated muon number $<N_{\mu }^{tun}>$ versus number of
electrons $<N_{e}>$

Figure 12. Differential spectrum of truncated muon number

Figure 13. Muon lateral distributions in comparison with CORSIKA and ARES
simulation results:

a) hall, b) tunnel

\end{document}